# Near shot-noise limited time-resolved circular dichroism pump-probe spectrometer

Valentyn Stadnytskyi 1), Gregory S. Orf 2), Robert E. Blankenship 2), Sergei Savikhin 1)*

[1]Department of Physics and Astronomy, Purdue University, West Lafayette, IN

[2]Photosynthetic Antenna Research Center, Departments of Biology and Chemistry, Washington University in St. Louis,

St. Louis, MO

*sergei@purdue.edu

We describe an optical near shot-noise limited time-resolved circular dichroism (TRCD) pump-probe spectrometer capable of reliably measuring μdeg circular dichroism signal with nanosecond time resolution. Such sensitivity is achieved through a modification of existing TRCD designs and introduction of a new data processing protocol that eliminates approximations that have caused substantial nonlinearities in past measurements and allows the measurement of absorption and CD transients simultaneously with a single pump pulse. The exceptional signal-to-noise ratio of the described setup makes TRCD technique applicable to a large range of non-biological and biological systems. The spectrometer was used to record, for the first time, weak TRCD kinetics associated with triplet state energy transfer in the photosynthetic Fenna-Matthews-Olson antenna pigment-protein complex.

**Introduction**

Circular dichroism (CD) spectroscopy is a quick and non-invasive tool for structural investigation of biological molecules and systems[1–3]. Time-resolved circular dichroism (TRCD) spectroscopy has been shown to have great potential for investigation of three-dimensional structural evolution during the reactions of biomolecules and chiral-chemical species[1,4]. However, the majority of nanosecond-microsecond TRCD experiments have so far been limited to species with strong transient CD signals on the order of >1 mdeg[5–28]. For example, transient CD signals measured for myoglobin[5] were on the order of ~10 mdeg with the signal to noise ratio of only ~5; transient signals measured for Ruthenium blue dimer[6] in the order of ~20 mdeg were measured with signal-to-noise ratio of ~20 (see Supplementary Information for more examples). Moreover, only few papers explicitly state noise levels of the proposed setups: 3.3 mdeg ($10^{-4}$ OD)[16] and 0.4 mdeg (or $1.3·10^{-5}$ OD)[5] with 30 minutes averaging. To best of our knowledge, none of the proposed setups was used to study systems with relatively weak transient CD changes below 1 mdeg with noise levels below 0.4 mdeg(~$10^{-5}$ OD). This work is the first realization of a nanosecond transient CD spectrometer capable of reliable measuring transient signals as low as 0.04 mdeg ($10^{-6}$ OD) in 60 seconds of integration time with nanosecond resolution and noise levels. One of the important classes of systems with low transient CD signals is photosynthetic proteins, where excitonic coupling between chromophores produces characteristic excitonic CD signature that carries rich information on the structure of the complex as well on the dynamics of the energy transfer process. For example, Fenna-Matthews-Olson(FMO) protein-pigment antenna complex is expected to have transient CD on the order of 0.5 - 0.1 mdeg, which is at or below the noise level that is published or can be inferred from the data presented in the literature so far (see examples in Supplementary Information) Hence, the improvement in the TRCD techniques and reduction of noise to the levels of shot-noise is important to expand the capabilities of TRCD techniques to this class of systems.

There are two widely used TRCD techniques: 1) absorption measurements utilizing modulation techniques, predominantly used in the earlier days on a slow (millisecond) time scale and later by femtosecond TRCD spectroscopy[6,16–21,29], and 2) ellipticity measurements, which gives the ability to retain nanosecond-microsecond time-resolutions while still measuring CD signals[5,10,12–15,18,21–28,30]. In the absorption measurements, a small difference in absorption of right and left circularly polarized light ($A_{CD}$) is recorded. The time-resolution in this case is limited by a modulation frequency not exceeding 84 kHz for photoelastic modulator, a requirement of a large fast-changing field for electro-optical modulators to reach higher modulation frequencies, and utilization of a time-delay (in the case of femtosecond resolution). The sensitivity of this method is generally low because changes in absorption of a sample upon excitation due to chirality are several orders of magnitude smaller than ordinary absorption changes. The detection of the small change on top of a large background is challenging. In contrast, ellipticity measurements can be background-free and the fraction of the chiral signal in the total probe beam intensity can reach 100 percent under certain experimental conditions. However, the ellipticity method is highly sensitive not only to CD, but also to other polarization effects such as optical rotation dispersion (ORD), circular birefringence (CB), linear dichroism (LD) and linear birefringence (LB)[9,31]. Fortunately, for the majority of nanosecond and millisecond TRCD experiments, effects of LD, LB and CB can be easily distinguished from CD by careful alignment and proper measurement protocol: LB and LD effects are nonexistent if the measured lifetimes are longer than the electronic state reorganization lifetime of the sample molecules, and the ORD effects can eliminated by measuring CD spectra at two different handednesses of the elliptically polarized light[31].

The first reported ellipsometric TRCD apparatus measured the kinetics of two different handednesses of elliptically polarized light separately[26], a similar design was used later by several groups[14,15,30]. However, in this approach the CD signal is a tiny difference between two sequential measurements that are separated in time, which introduces the additional challenge of measuring a weak chiral signal on top of a much stronger achiral signal, with the latter being susceptible to pump and probe light fluctuations and to sample degradation. In addition, due to mathematical approximations, the precision of the reconstructed CD depended on the parameters of the experimental setup. Hence, the previous CD measurements often had a more qualitative than a quantitative description of the sample of interest and could be applied only to samples with strong CD signals.

In this work, we present an improved TRCD spectrometer design with sensitivity approaching the shot-noise limit. This design eliminates drawbacks of previous setups, such as mechanical rotations of stress plate (e.g. [26]), Babine-Soleil compensators (e.g. [22]) and CD signal reconstruction error due to approximation (e.g. [26]), allowing precise measurement of TRCD kinetics, which are not affected by fluctuations in pump and probe light intensities, sample degradation, and measured CD signals. This is achieved by the extensive redesign of the detection system and development of a new signal reconstruction procedure. The analysis applied in this study reconstructs CD signals correctly under all experimental setup parameters and sample properties. Moreover, the described setup measures ordinary absorption (or optical density, OD) changes along with the changes in CD simultaneously with a single pump pulse. The sensitivity of the proposed setup is approaching shot noise and transient changes in CD as small as 20 µdeg (e.g. 0.15 cm$^{-1}$ M$^{-1}$ for FMO protein) can be reliably measured in a few minutes of integration time. To demonstrate the capability of the system, it was used to measure the wavelength-dependent dynamics of the near-IR CD signal of the bound bacteriochlorophyll pigments in the photosynthetic Fenna-Matthews-Olson pigment-protein antenna complex, where CD response upon laser excitation is dominated by excitonic sub-mdeg effects.

**Analytical framework**

The functional block diagram of the TRCD spectrometer is shown in Figure 1A. The design resembles that proposed by [26,32] and used by several other groups[10,14,15,31], except the presence of two additional photodetectors for recording the reference beam intensity (PD$_{ref}$) and full intensity of the transmitted beam (PD$_{full}$). The probe beam is passed through a linear polarizer (P$_1$) followed by a stress plate (OR) that introduces slight optical retardation with retardation +δ or -δ, depending on the orientation of the stress plate. Upon passing the sample, the intensity and retardation of the probe beam changes, reflecting absorption and CD properties of the sample. The transmitted beam intensity is measured by PD$_{full}$ and the retardation is analyzed through a second polarizer P$_2$, which is crossed at 90° with respect to P$_1$, by photodetector PD$_{perp}$. The signals expected in such a setup can be conveniently analyzed using Jones calculus [14,15,26,31,33,34].

The original setup used by Lewis et al.[26] had only one photodetector, $PD_{perp}$, and to extract the CD signal two separate measurements were performed with two different orthogonal orientations of the stress plate, one producing retardation of +δ, and another -δ, where retardation is defined conventionally as $\delta^2 = (I_{perp} - I_{bg})/(4I_{par})$, where $I_{perp}$ and $I_{par}$ are intensities along minor (measured by $PD_{perp}$) and major (measured by $PD_{full}$) axis of the elliptically polarized light and $I_{bg}$ is background signal in case of the crossed polarizers. Using Jones calculus it was shown[26], that in the limit of small retardation and small CD signal, the $A_{CD}$ absorption can be expressed as:

$$A_{CD} = \frac{\delta}{2.3} \frac{I_R - I_L}{I_R + I_L} \quad\quad 1$$

where $I_R$ and $I_L$ are time-dependent signals measured by $PD_{perp}$ for two different orientations of the stress plate, +δ and -δ, respectively. $A_{CD}$ is defined here as the difference between the absorption of left circularly polarized light ($A_L$) and right circularly polarized light ($A_R$): $A_{CD}=A_L - A_R$. This quantity can be converted to mdeg as follows: 1 mdeg = 32.982·$A_{CD}$. The above scheme measures absolute $A_{CD}$, with the transient $\Delta A_{CD}$ due to sample excitation being on top of the steady state $A_{CD}$. Using the above scheme, Lewis et al.[26] succeeded in detecting the excitation-induced changes in $A_{CD}$ associated with photolysis of (carbonmonoxy)myoglobin with a signal on the order of $\Delta A_{CD} \sim 6 \times 10^{-4}$ (20 mdeg) with a signal-to-noise ratio of about 5 to 10. Measuring weaker signals using this approach is problematic; the $A_{CD}$ signal of Lewis et. al.[26] is only a small difference (~1%) between relatively large separately measured $I_R$ and $I_L$, which can lead to a significant error in reconstituted $A_{CD}$ profile due to small pulse-to-pulse fluctuations in probe or pump light intensities, or due to even tiny absorption changes caused by sample degradation between the pulses.

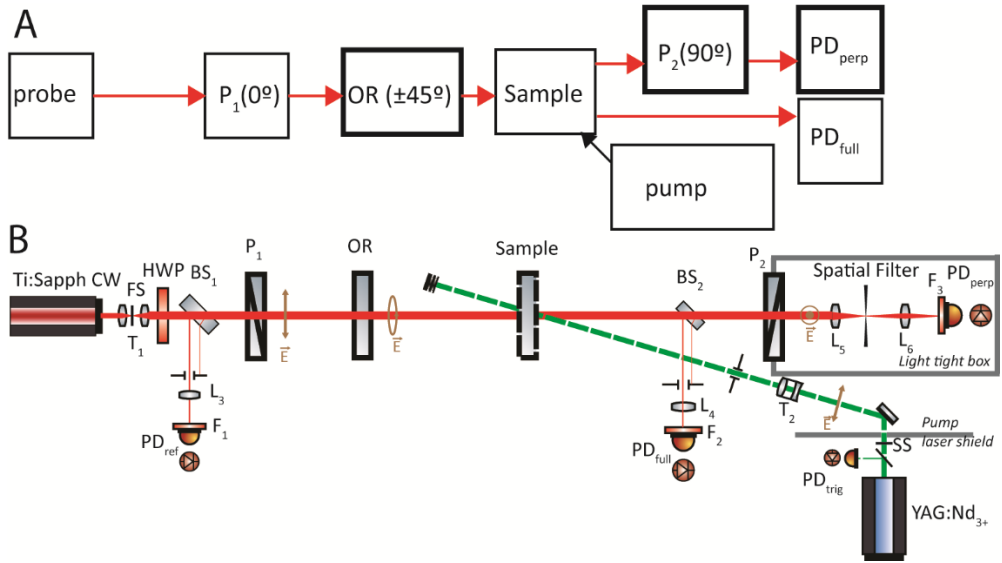

Figure 1: A —Block diagram of Jones matrices representing each element of the optical setup. B — Optical scheme of the TRCD pump-probe spectrometer. Probe — probe light source (CW Ti:Sapphire laser); $P_1$ — polarizer; OR — optical retarder; $P_2$ — analyzer; $PD_{perp}$, $PD_{full}$, $PD_{ref}$ —photodetectors; pump – pump pulse source (Nd:YAG pulsed laser); FS and SS – fast and slow shutters; $BS_{1,2}$ – beam splitters; L – lenses; HWP – Half-wave plate; F – filters; $T_1$ and $T_2$ –Kepler and Galilean telescopes, respectively. In Panel A orientations are given in brackets relative to the y-axis.

The degrading effect of the probe beam fluctuations in our design of TRCD spectrometer is addressed by the addition of photodetector $PD_{ref}$ (Figure 1) that monitors the probe beam intensity ($I_{ref}$) before the sample. Using $I_R/I_{ref}$ and $I_L/I_{ref}$ in place of $I_R$ and $I_L$ in Equation 1 will diminish the effect of pulse-to-pulse fluctuations in the probe light; but as we learned experimentally, that alone is not sufficient to achieve shot-noise limited detection of weak $\Delta A_{CD}$ signals.

The addition of one more photodetector, $PD_{full}$, to monitor the intensity of probe beam after the sample enables measurement of both ordinary $\Delta A$ and transient background-free $\Delta A_{CD}$ signals with a *single* pump pulse, eliminating the necessity to perform two separate measurements with two orientations of the stress plate and two separate pump pulses. Our numerical analysis shows that, in the limit of small signals, the transient $\Delta A_{CD}$ signal in this case can be expressed as (see Supplementary Information for details):

$$\Delta A_{CD} = \frac{4}{2.3\delta} \left\{ \left[ \frac{I_{perp}}{I_{full}} \right]^{with\ pump} - \left[ \frac{I_{perp}}{I_{full}} \right]^{without\ pump} \right\} \qquad 2$$

Note that in this scheme the $\Delta A_{CD}$ signal is *background-free*, i.e. it is automatically isolated from the steady state $A_{CD}$. Since the ratio $I_{perp}/I_{full} \sim \delta^2$, the relative changes in $I_{perp}$ due to the presence of pump beam are $\sim 1/\delta$ times larger than the respective $\Delta A_{CD}$, i.e., the ellipsometric scheme dramatically amplifies the small $\Delta A_{CD}$ signals. Note also that $(I_{perp}/I_{full})^{without\ pump}$ is essentially linear; any deviations in $(I_{perp}/I_{full})^{with\ pump}$ in response to the pump indicate the presence of non-zero $\Delta A_{CD}$, allowing for easy real-time visualization of the signal on an oscilloscope capable of plotting the ratio of two input channels. The ordinary $\Delta A$ is calculated as usual:

$$\Delta A = -\log_{10} \frac{\left[ I_{full}/I_{ref} \right]^{with\ pump}}{\left[ I_{full}/I_{ref} \right]^{without\ pump}} \qquad 3$$

Equations (1) and (2) are approximations where only a higher term is left and all smaller are equated to zero (derivation of the equation (1) is described in details elsewhere[26]). This approximation works well only under conditions when a leading term is much larger than other terms and can produce large error if these conditions are not met which is always true with approximations. However, further analysis of our TRCD setup (Fig. 1A) using Jones calculus shows that if all three signals ($I_{full}$, $I_{perp}$, and $I_{ref}$) are measured simultaneously then the resulting system of four equations for ordinary absorption ($A$, $\Delta A$, $A_{CD}$ and $\Delta A_{CD}$) can be solved exactly (see Supplementary Information for details):

$$\begin{cases} I_{perp}^{without\ pump} = \frac{1}{4}\left[4k(k-\rho)\sin^2(\tfrac{\delta}{2}) + \rho(2k-\rho)\sin(\delta) + \rho^2\right]I_{ref} \\ I_{full}^{without\ pump} = kk'I_{ref} \end{cases} \Biggr\} without\ pump$$

$$\begin{cases} I_{perp}^{with\ pump} = \frac{1}{4}\left[4k(k-\rho)\sin^2(\tfrac{\delta}{2}) + \rho(2k-\rho)\sin(\delta) + \rho^2\right]I_{ref} \\ I_{full}^{with\ pump} = kk'I_{ref} \end{cases} \Biggr\} with\ pump \qquad 4$$

$$k = 10^{-A_R/2}; k' = 10^{-A_L/2}; \rho = k - k'$$

Here $A_R$ and $A_L$ are the absorbances of right and left-hand circular polarized light, $\delta$ is the retardation of the stress plate (OR), and $I_{ref}$, $I_{perp}$, $I_{full}$ are light intensities measured with $PD_{ref}$, $PD_{perp}$ and $PD_{full}$ photodiodes, respectively. In the first two equations, $k$ and $\rho$ are defined for the case when pump light was absent, and in the second two these parameters correspond to the case when pump light was present.

The analytical solution of Eq. 4 for $\Delta A$ and $\Delta A_{CD}$ provided by Mathematica is too bulky, and thus in our data analysis we used a more compact Newton-Gauss numerical method to solve the system of the above equations. While equations 2 and 3 are very convenient for quick real-time processing of the data for visualization, the exact numerical solution of the system of equations needs to be done since approximate equations do not quantitatively reconstruct measured CD signals as was also noted in[26]. Different computational approaches of reconstruction of CD signal using measured quantities were compared using a numerical simulation (Figure 2). The TRCD setup in simulation was described with a set of Jones matrices for each optical element and a Jones vector as an input light intensity and polarization, similar to the original work by Lewis at al.[26]. The simulations take input light intensity ($I_{ref}$), absorbance (k and k'), circular dichroism ($\rho$), and retardation ($\delta$) of the stress plate to compute intensities after an ideal polarizer and ideal photodiode. Further, the computed intensities are used in Eqs. 1, 3, and 4 to compute circular dichroism and compare it with the input parameters. The simulation output results for the same hypothetical data set as input are shown in the Figure 2.

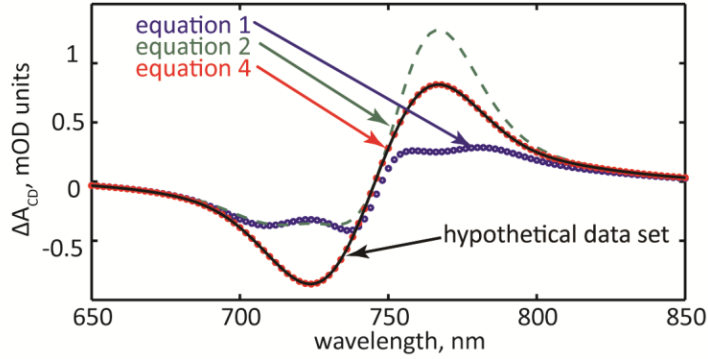

Figure 2: Comparison of the hypothetical CD data set and reconstructed CD signals with three different methods assuming the hypothetical data set as an input of TRCD spectrometer: Blue circles – eq. 1; Green dashed line – eq. 2; Red circles – eq. 4 and Black solid line – hypothetical data set. Note that the asymmetry in reconstruction with Eq. 2 is due to the sign of the used delta ($\delta=0.016$).

**Experimental setup configuration**

The full optical diagram of the TRCD experimental setup used in this work is shown in Figure 1B. A probe light was provided by a home-built CW Ti:Sapphire laser with output power up to 1.1 W tunable over 700-930 nm, which was pumped with the 5 W CW 532-nm output of a Millennia Ev5 laser (Spectra-Physics Inc.). A Kepler telescope ($T_1$) was used to enlarge the diameter of the probe beam to ~3 mm to decrease the effect of non-homogeneity in stress distribution in a stress plate (OR) and reduce exposure of the sample. The fast-mechanical shutter (Vincent Associates VS35 and VCM-D1) opens only for ~1 ms during the measurement to reduce the exposure of the sample to the probe light. The probe beam was then passed through the half-wave plate (HWP) (ThorLabs WPH10E-780) to compensate for any rotations in probe light polarization after reflection of the probe beam guiding mirrors (not shown). A thick fused silica glass was used as a beam splitter $BS_1$ to spatially separate the light beams reflected from two sides and avoid interference effects that can be severe in the case of the coherent laser beam. The reflected reference beam was measured by a photodiode ($PD_{ref}$). The transmitted beam was passed through a polarizer $P_1$, which ensured high-contrast vertical polarization of the probe light. The polarized light was then passed through an optical retarder (OR) comprised of a 6.35-mm-thick fused silica precision wedged window (Esco Optics). The window was mechanically compressed along a line at 45° to the polarization of the input beam introducing slight retardation of the linearly polarized light and making it elliptically polarized. It is essential to use a wedged plate to avoid multiple reflections in the plate that accumulate different retardation and later interfere with each other causing a significant deviation in ellipticity of the probe light. The elliptically polarized light passed through a sample of interest and was then split into two by another thick fused silica window $BS_2$ to eliminate interference effects. The reflected beam was detected by a photodiode ($PD_{full}$). The transmitted beam passed through a crossed analyzing polarizer $P_2$, a spatial filter to reduce scattered probe light and fluorescence effects, and was finally detected with a third photodiode ($PD_{perp}$). Note that in the absence of the stress plate and a sample, light to $PD_{perp}$ must be fully blocked by the two crossed polarizers, and it is essential that both polarizers $P_1$ and $P_2$ have the highest possible extinction ratio. In the described setup, both polarizers were of the Glan-Thompson type with an extinction coefficient better than $5 \times 10^{-6}$ (Artiflex engineering). All photodiodes had a 2.3-ns response time (PDA10A, Thorlabs). The long-pass filters ($F_1$ - $F_3$) are used to block scattered pump light and let the probe light through.

A 5-ns pump pulse was provided by tunable optical parametric oscillator pumped with a pulsed YAG:$Nd^{3+}$ laser (Ekspla NT342B-10SH-WW). The diameter of the pump pulse beam was set to be slightly larger than that of the probe beam in the sample to ensure excitation throughout the volume of the probed sample. The polarization of the probe pulse was vertical to diminish scattering of pump pulse onto the $PD_{perp}$. The repetition rate of the pump pulses was set to 0.3 Hz.

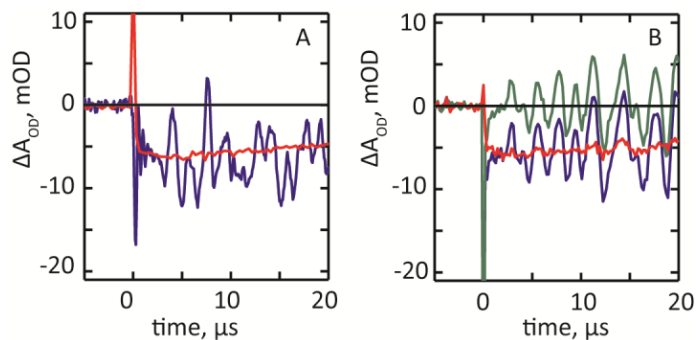

Figure 3: $\Delta A_{OD}$ in the perpendicular channel (CD sensitive) measured for FMO complex in a standard and home-built cells after excitation at 600 nm. (A) Blue: Acoustic oscillations in a standard rectangular 1-mm-thick Starna cell give rise to oscillating $\Delta A_{OD}$ probed for FMO at 825 nm. Red: Measurement of $\Delta A_{OD}$ in a special home-built cell is oscillation free. (B) $\Delta A_{OD}$ probed for FMO in a Starna cell at 825 nm (blue) and 850 nm (green). Since the acoustic oscillation effect is almost independent of the probe wavelength and there is no signal from FMO at 850 nm, subtracting the 850-nm kinetics from the 825 nm cancels oscillations in FMO kinetics at 825 nm (red curve).

A home-built cell with strain-free windows and temperature control was used for time-resolved measurements. The cell consists of two strain-free round glass coverslips (0.17-mm-thick, 22-mm-diameter, HR3-231, Hampton Research) separated by a 1-mm-thick rubber spacer and pressed together in a home-made metal holder. For temperature control, the temperature of the metal holder was stabilized by the water bath (ThermoForma 003-8818), and a K-type thermocouple was used to measure the temperature inside the cell. The use of thin round windows, rubber spacer, and a lower temperature were absolutely necessary to reduce the acoustic wave effects created by a slight abrupt rise of temperature (<0.1 °C) in response to the pump pulse; this acoustic shock initiates standing strain waves in a standard rectangular sample cell leading to oscillations in $\Delta A_{OD}$ that are significantly stronger than the shot-noise limit of the described setup (see Figure 3A). The round shape minimizes the anisotropic nature of these waves, thin windows minimize the induced CD amplitude while setting the temperature of a sample diluted in water to 4 °C minimizes the thermal expansion of water and the magnitude of the temperature-induced acoustic shock wave. Alternatively, a standard rectangular optical cell can be used if there is a known wavelength in the sample at which CD response of the sample to pump pulse is zero, provided that this wavelength is close to the wavelength where the sample response is measured. Since the acoustic TRCD response is highly reproducible and weakly dependent on the probe wavelength, the signal measured at this wavelength can be subtracted from the signal of interest to eliminate oscillations from the signal (Figure 3B).

The signals from all photodiodes were digitized in parallel by a four channel Tektronix TDS7154B oscilloscope with 1.6-ns resolution. All measurements were done with a vertical resolution of 20 mV/div to reduce the digital noise of the oscilloscope; the signals from the photodiodes were attenuated accordingly, and constant background subtracted whenever necessary. The oscilloscope was connected to a computer and data for each kinetic trace after each laser shot was extracted and saved for further analysis. Both $\Delta A$ and $\Delta A_{CD}$ were computed by solving the system of four nonlinear equations (Eq. 4) by the Newton-Gauss method for every point in time.

**Time-resolved circular dichroism measurements**

To confirm the exceptional sensitivity of the proposed setup, we have measured TRCD spectra of the light-harvesting Fenna-Matthews-Olson (FMO) pigment-protein complex from the photosynthetic green sulfur bacteria[35]. The Fenna-Matthews-Olson complex is a protein homotrimer in which each monomer binds eight light-absorbing bacteriochlorophyll *a* (BChl *a*) molecules. Its primary function is to channel electronic light excitation from a larger light-harvesting antenna, the chlorosome, to the membrane-bound photosynthetic reaction center. Note that the FMO complex was the first photosynthetic pigment-protein complex for which a three-dimensional x-ray crystal structure was determined, which led to it becoming a model system in photosynthesis research[36]. While a free BChl *a* molecule is planar and does not exhibit any significant CD, the close spacing and specific orientation of the eight BChls results in strong intermolecular dipole-dipole interactions. This leads to excitonic delocalization of the excited states over multiple pigments[37]. The excitonic character of excitations in this system leads to remarkable effects including quantum coherence[38,39]. Despite significant research over the past forty years, there are still ~10 different Hamiltonians still in use to describe its optical properties[40–47]. Due to excitation delocalization over a space of multiple pigments, this complex also exhibits a rich CD spectrum spreading over the entire excitonic $Q_y$ band of BChls between 770-830 nm[42,48,49]. This CD spectrum is, in fact, more sensitive to the details of the structure than a conventional absorption spectrum and, as our exciton modeling shows, can readily distinguish between different Hamiltonians proposed to describe the properties of FMO. However, the TRCD signals in FMO are expected to be in the order of ~$10^{-5}$ (300 µdeg), which is smaller than signals measured in the previously described ellipsometry based TRCD spectrometers. This new TRCD spectrometer was built to enable such kind of measurements.

The preparation and purification procedure of FMO sample is described elsewhere[50]. The protein sample was diluted in 3-(Cyclohexylamino)-1-propanesulfonic acid (CAPS) buffer and had an absorbance of ~0.6 at the 808 nm absorption band maximum[51–53]. Upon light excitation, a triplet excited state is formed with about 11% quantum yield, which leads to transient changes in both absorption and CD properties of this system[42,50,53]. Fig. 4A shows transient signals probed at 820 nm with $PD_{full}$ and $PD_{perp}$ after exciting FMO sample at 600 nm and treated as ordinary absorption changes, *i.e.* computed according to Eq. 3 ($\Delta A_{full}$ and $\Delta A_{perp}$, respectively). Note that $\Delta A_{full}=\Delta A$ while $\Delta A_{perp}$ will deviate from that only in the case of a nonzero $\Delta A_{CD}$. The two pairs of these profiles are shown that were measured with two orthogonal orientations of the stress plate corresponding to retardation $+\delta$ and $-\delta$. There is a clear difference in both the amplitudes and shapes of $\Delta A_{full}$ and $\Delta A_{perp}$, which indicates the presence of a nonzero $\Delta A_{CD}$. Also, the $\Delta A_{perp}$ measured with stress plate in $+\delta$ and $-\delta$ positions are not the same, indicating that there is a true $\Delta A_{CD}$ signal, since in the case of ORD and LD the $\Delta A_{perp}$ signal would be independent of the sign of $\delta$. Thus, all measurements were conducted at both $+\delta$ and $-\delta$ positions of the stress plate to ensure that ORD and LD effects are not present in FMO. The magnitude of the difference between $\Delta A_{full}$ and $\Delta A_{perp}$ is directly related to our ability to extract $\Delta A_{CD}$ from the experiment, and that difference will decrease rapidly with increase in the magnitude of retardation $\delta$.

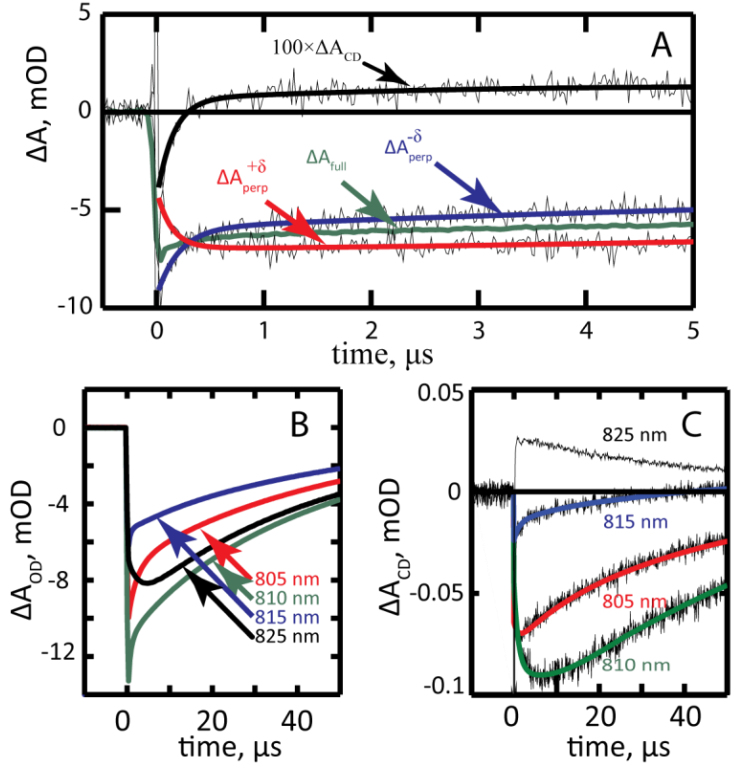

Figure 4: Time-resolved OD and CD kinetic traces for FMO excited at 600 nm. (A) $\Delta A_{per}^{+\delta}$ (red) and $\Delta A_{per}^{-\delta}$ (blue) are absorption differences measured with PD$_{perp}$ for two opposite retardations δ and -δ. $\Delta A_{full}=\Delta A$ (green) is an ordinary absorption difference measured with PD$_{full}$ that is independent of δ. $\Delta A_{CD}$ (black) is a respective transient CD signal calculated using Eq. 4 (B, C) – $\Delta A$ and $\Delta A_{CD}$ probed at 825, 815, 810, and 805 nm: colored curves are multiexponential fits, thin gray lines are measured profile, except the 825 nm where black is an actual measured profile and multiexponential fit is not show because the noise will not be visible behind the fit line. The noise in Panel B is smaller than the thickness of the fit lines.

Figure 4B and 4C present the results of simultaneous measurement of ordinary $\Delta A$ and $\Delta A_{CD}$, respectively, for the FMO complex at 4 °C after excitation at 600 nm. The excitation pulse energy was 3 mJ and the probe beam intensity was 250 mW with δ=0.02. All signals were averaged over ~300 pump pulses except the 825nm kinetics which was averages over 3000 pump pulses. The signals measured are similar to the ones shown in Figure 4A, but the data were processed to yield correct $\Delta A$ and $\Delta A_{CD}$ signals using a precise solution of Eq. 4 as described in the previous section. Signals probed at several wavelengths are shown here with the sole purpose of demonstrating the high sensitivity of the spectrometer; detailed analysis of the data for the purpose of further analyzing the optical characteristics of the FMO complex will be published elsewhere. The exponential fits to $\Delta A$ and $\Delta A_{CD}$ kinetics reveal that both can be described with four components. Three of the observed lifetimes were assigned earlier to singlet exciton absorption evolution due to triplet excited state energy transfer between different BChl *a* pigments within the same FMO complex (1 μs, 11 μs) and overall triplet excited state decay to the ground state (55 μs)[53]. The additional 100-ns lifetime has not been previously observed, and its decay-associated spectrum and analysis will be published elsewhere. As predicted by exciton simulations (also to be published elsewhere), the amplitudes and signs of these components are not the same for $\Delta A$ and $\Delta A_{CD}$ profiles, reflecting different dependencies of absorption and CD spectra on the structural and energetic parameters of an excitonic system, with the $\Delta A_{CD}$ signals being more sensitive to slight changes in the model than conventional $\Delta A$.

Note that ORD and/or LD contributions to the $\Delta A_{CD}$ signals computed using Eq. 4 would change sign when swapping retardation between $+\delta$ and $-\delta$ (by turning the stress plate OR by 90°). Thus, their contribution can be revealed and eliminated by performing two consequent measurements with two orientations of the stress plate (+45° and -45° in respect to input polarization), and computing the true signal as $\Delta A_{CD} = \left(\Delta A_{CD}^{+\delta} + \Delta A_{CD}^{-\delta}\right)/2$. We found that in the case of FMO there was no detectable ORD/LD contribution.

The measured $\Delta A_{CD}$ kinetic amplitudes in Figure 4C are on the order of $10^{-5}$, which is at or below the noise level of the previously proposed TRCD spectrometers (see Supplementary Information), indicating that such measurements would be impossible to perform using those setups. These measurements have become possible through our instrumental design and analytical framework, giving us the necessary sensitivity resulting in a signal-to-noise ratio of ~100. As a bonus, this setup also provides exceptional sensitivity in detecting ordinary $\Delta A$ kinetics; the noise spread in Figure 4B is smaller than the line thickness.

The absolute noise level in the measured $\Delta A_{CD}$ profile at 825nm in Figure 4C is ~$7.8 \times 10^{-7}$ (calculated as rms). This level is comparable to that of the shot-noise ($3.37 \times 10^{-7}$) predicted on the basis of the number of photons in the probe light (See supplementary Information), meaning that this system's sensitivity is at the best possible level; further improvement is impossible without a substantial increase in the probe light intensity.

In conclusion, the TRCD setup developed in this work offers for the first time a near shot-noise limited performance. It was successfully used to observe weak time-resolved circular dichroism changes in a photosynthetic light-harvesting protein for the first time. The proposed design can be readily adapted to cover different wavelength ranges and applied to a broad range of biological and non-biological samples.

**Acknowledgements**

Spectroscopic and computational studies by V.S. were supported by the Division of Chemical Sciences, Geosciences, and Biosciences, Office of Basic Energy Sciences of the U.S. Department of Energy through Grant DE-FG02-09ER16084 to S.S. Purification and collection of the FMO protein complex by G.S.O. were supported by the Photosynthetic Antenna Research Center (PARC), an Energy Frontier Research Center funded by the U.S. Department of Energy, Office of Science, Office of Basic Energy Sciences under award number DE-SC0001035 to R.E.B.